\documentclass[usenatbib,onecolumn]{mnras}

\usepackage{newtxtext,newtxmath}
\usepackage[normalem]{ulem}

\usepackage[T1]{fontenc}

\DeclareRobustCommand{\VAN}[3]{#2}
\let\VANthebibliography\thebibliography
\def\thebibliography{\DeclareRobustCommand{\VAN}[3]{##3}\VANthebibliography}

\usepackage{CJKutf8}

\usepackage{amsmath}
\newcommand{\vecq}{\mathbfit{q}}
\newcommand{\vecr}{\mathbfit{r}}
\newcommand{\veck}{\mathbfit{k}}
\newcommand{\vecx}{\mathbfit{x}}
\newcommand{\vecv}{\mathbfit{v}}
\newcommand{\vecd}{\mathbfit{d}}
\newcommand{\rmi}{{\rm i}}
\usepackage{color}  
\usepackage{mathrsfs}
\usepackage{xcolor,cancel,ulem}

\usepackage{graphicx}	
\usepackage{amsmath}	

\title[Velocity correlation function]{The effects of non-linearity on the growth rate constraint from velocity correlation functions}

\author[M. Tonegawa et al.]{
Motonari Tonegawa,$^{1}$\thanks{E-mail: motonari. tonegawa@apctp.org}
Stephen Appleby,$^{1,2}$
Changbom Park,$^{3}$
Sungwook E. Hong,$^{4,5}$
and Juhan Kim$^{6}$
\\
$^{1}$Asia Pacific Center for Theoretical Physics, Pohang, 37673, Korea\\
$^{2}$Department of Physics, POSTECH, Pohang, 37673, Korea\\
$^{3}$School of Physics, Korea Institute for Advanced Study, 85
Hoegiro, Dongdaemun-gu, Seoul, 02455, Korea\\
$^{4}$Korea Astronomy and Space Science Institute,
776 Daedeok-daero, Yuseong-gu, Daejeon, 34055, Republic of Korea\\
$^{5}$Astronomy Campus, University of Science and Technology,
776 Daedeok-daero, Yuseong-gu, Daejeon, 34055, Republic of Korea\\
$^{6}$Center for Advanced Computation, Korea Institute for Advanced Study, 85 Hoegiro, Dongdaemun-gu, Seoul, 02455, Korea
}

\date{Accepted XXX. Received YYY; in original form ZZZ}

\pubyear{2024}

\begin{document}
\label{firstpage}
\pagerange{\pageref{firstpage}--\pageref{lastpage}}
\maketitle

\begin{abstract}
The two-point statistics of the cosmic velocity field, measured from galaxy peculiar velocity (PV) surveys, can be used as a dynamical probe to constrain the growth rate of large-scale structures in the universe. 
Most works use the statistics on scales down to a few tens of Megaparsecs, while using a theoretical template based on the linear theory. In addition, while the cosmic velocity is volume-weighted, the observable line-of-sight velocity two-point correlation is density-weighted, as sampled by galaxies, and therefore the density-velocity correlation term also contributes, which has often been neglected.
These effects are fourth order in powers of the linear density fluctuation $\delta_{\rm L}^4$, compared to $\delta_{\rm L}^2$ of the linear velocity correlation function, and have the opposite sign. We present these terms up to $\delta_{\rm L}^4$ in real space based on the standard perturbation theory, and investigate the effect of non-linearity and the density-velocity contribution on the inferred growth rate $f\sigma_8$, using $N$-body simulations. We find that for a next-generation PV survey of volume $\sim {\cal O}(500 \, h^{-1} \, {\rm Mpc})^3$, these effects amount to a shift of $f\sigma_8$ by $\sim 10$ per cent and is comparable to the forecasted statistical error when the minimum scale used for parameter estimation is $r_{\rm min} = 20 \, h^{-1} \, {\rm Mpc}$. 
\end{abstract}




\section{Introduction} \label{sec:intro}
The velocity field of the universe is an important cosmological probe, due to its capability of constraining the dynamical evolution of the universe \citep{Ferreira:1999}. It can be used to discriminate between cosmological models and gravity theories that otherwise may have indistinguishable observables related to the background geometry, such as BAO, CMB, and type Ia supernovae \citep{Eisenstein:2005,Dunkley:2009,Aghanim:2018,Bautista:2018,Abbott:2019}. Therefore, many upcoming cosmological surveys include measuring the cosmic velocity field as one of their key science objectives (DESI: \citet{Aghamousa:2016}; Euclid: \citet{Amendola:2018}).

There are two kinds of observations to probe the cosmic velocity field. One is redshift space distortions \citep[RSD;][]{Kaiser:1987,Hamilton:1998}, as probed by spectroscopic data. Redshifts measured using spectral features of galaxies are shifted by the Doppler effect caused by the peculiar velocity (PV) of galaxies. This velocity field is 
 sourced by the gravitational potential of the matter distribution and coherent on large scales, causing a compression of the two-point statistics along the line of sight on large scales. The degree of anisotropy is sensitive to the growth rate parameter $f\sigma_8$. There have been many successful detections of RSD \citep{Guzzo:2008,Blake:2011,Samushia:2012,10.1093/mnras/stu1391,Okumura:2016}, placing a constraint on the growth rate of the large-scale structure.
The second is inference of the velocity field from distance measurements. Certain classes of galaxies are known to have a tight correlation between stellar kinematics and luminosity \citep{Faber:1976,Tully:1977}. These relations allow us to deduce the luminosity of galaxies, which in turn can be translated into true distances. Combined with spectroscopic redshifts, one can extract PVs, because the observed redshift comprises a combination of the Hubble flow and PV.
Numerous theoretical studies relating to PV statistics have been carried out \citep{Park:2000,Koda:2013eya,Saito:2014,Okumura:2016,Castorina:2019hyr}, and measurements with existing data has been undertaken \citep{Davis:1982gc,Park:2005bu,Johnson:2014kaa,Howlett:2017asq,Adams:2017val}. With increasingly large catalogs becoming available, PV draws attention as a promising way to constrain cosmological models \citep{Howlett:2016urc,Dupuy:2019,Turner:2021,Tully:2023}.

The observable of a PV survey data can be either in configuration or Fourier space. In Fourier space, at linear scales the Fourier modes are independent of each other and thus we can expect a clean measurement. However, in practice, the survey mask induces artificial mode couplings and this effect should be carefully taken into account \citep{Park:2000,Park:2005bu,Howlett:2017,Qin:2018}. Also, it has been recently argued that the non-linear effects and RSD starts to be effective on relatively large scales \citep{Dam:2021}. Another fundamental problem is that, in underdense or void regions, the cosmic velocity cannot be sampled by tracer galaxies, even though there exists a velocity flow in the dark matter. Modelling the field using interpolation schemes in such regions could result in a biased measurement. For this reason, it is more natural to use the momentum power spectrum, $p=(1+\delta)v_{||}$, where the value approaches zero in such regions \citep{Park:2000}.

In configuration space, the quantity of interest is the two-point correlation function of the line-of-sight component of the peculiar velocity as a function of pair separation $r$. This quantity is intuitively simpler to interpret because it is relatively easy to subtract the mask convolution effects.
If the size of the sample is not large, as is the case for PV surveys, the calculation is not overly intensive, only requiring us to count the pairs as a function of separation. Also, it is easy to incorporate the effect of survey mask. \citet{Gorski:1989} defined quantities $\psi_1$ and $\psi_2$, which can be readily used for any survey geometry, because it includes the information of the survey geometry via the quantities $A(r)$ and $B(r)$, as described in the following section. Furthermore, the correlation function is less sensitive to the shot-noise like component \citep{Cohn:2006} than the momentum power spectrum \citep{Howlett:2017, Howlett:2019bky, Appleby:2023}.
The utility of these quantities is being studied by many authors \citep{Wang:2018,Dupuy:2019,Turner:2021,Blake:2024}.

The underlying premise to use PV data to constrain cosmology is that the velocity divergence field $\theta(\vecx)$ is proportional to the overdensity $\delta(\vecx)=(\rho(\vecx)-\bar{\rho})/\bar{\rho}$ in Fourier space: $\tilde{v}(\veck)=-{\rm i}(aHf)\frac{\veck}{k^2}\tilde{\theta}(\veck)=-{\rm i}(aHf)\frac{\veck}{k^2}\tilde{\delta}(\veck)$. 
This leads to the relation between the velocity power spectrum $P_{vv}$ with $v = {|\boldsymbol{v}|}$ being the absolute value of the cosmic velocity field, and the linear matter density power spectrum $P_{\rm L}(k)$; $P_{vv}(k)\propto f^2 P_{\rm L}(k)$ in the linear regime\footnote{The amplitude is also sensitive to the normalization of the matter power spectrum $\sigma_8$, thus the actual parameter constrained is $f\sigma_8$.}. 
Thus the quantity of interest is the amplitude of the velocity two-point statistics. It is often assumed that the velocity correlation can be modeled using linear theory. Linear theory is valid on large scales, and as a factor of $1/k^{2}$ is present in the conversion between the velocity power spectrum and correlation function, somewhat small scales can be used in configuration space analysis. However, the modest volume of data currently available requires us to go down to small scales to obtain useful cosmological constraints. It has been found that non-linear effects in the velocity statistics start to become important at $k\sim 0.1 h \, {\rm Mpc}^{-1}$ \citep{Park:2000,Park:2005bu,Okumura:2014,Howlett:2017,Dam:2021,Appleby:2023}. This corresponds to a non-linearity scale $r_{\rm NL} \sim 2 \pi/0.1 \sim 60 \, h^{-1} \, {\rm Mpc}$. Transforming between Fourier and configuration space requires integration over Fourier modes, so the additional $k^{-2}$ factor is likely to mitigate non-linear effects in the velocity correlation function, unlike the density correlation function. On the other hand, the integration delivers the non-linear effects to large scales, unlike the Fourier space analysis, where the independence of different $k$-modes grants clean measurements on large scales. Therefore,
it will be necessary to test how non-linear effects will impact cosmological inference from the velocity correlation function, prior to the application of the theory to data, especially if we intend to use scales considerably below the limit $r_{\rm NL}$, which has been the case in the literature,

Another issue which has not been fully taken into account is that the quantity we measure is density-weighted rather than volume-weighted. This comes from the fact that we can only observe the velocity of galaxies, and galaxies trace the density field. Thus the velocity statistics that we obtain is weighted according to the density field, and we cannot measure the pure velocity of the dark matter flow within a given volume. This generates an additional contribution to the correlation function, which is the density-velocity correlation and scales as $O(\delta_{\rm L}^4)$, where $\delta_{\rm L}$ is the linear theory density fluctuation. This was pointed out by several authors \citep{Sheth:2008,Okumura:2014}, but it is often neglected because on large scales its effect is subdominant.

Bearing these points in mind, we study additional effects beyond the linear theory velocity-velocity correlation, up to order $\delta_{\rm L}^4$. We use two $N$-body simulations, one dark matter only and the other mock galaxies, to measure the velocity two-point correlation function. By fitting them using both linear and non-linear theory, we examine how inferred values of $f\sigma_8$ are shifted by the non-linearity, and compare the shift to statistical errors. This will inform us of the suitability of linear theory, as a function of the minimum scale used. 

In this work we exclusively study the correlation function in real space, whereas the analysis of actual observations is usually performed in  redshift space. We restrict ourselves to real space because our objective is to precisely quantify the difference in using linear and higher order perturbation theory on cosmological parameter estimation. In redshift space, the Finger-of-God effect -- an intrinsically non-perturbative phenomenon -- causes further damping in the amplitude and disentangling its impact on the correlation function compared to perturbative contributions introduces an additional layer of phenomenological modelling. Measuring the velocity field two-point statistics in real space may be a preferred option in the future, although this approach has its own problems \citep{Park:2005bu,Koda:2013eya}. 

The contents of the paper is as follows. In section \ref{sec:theory}, we review the theoretical aspects of the velocity correlation function. In section \ref{sec:floats}, we explain the data used. In section \ref{sec:fitting}, we present the fitting prescription utilised and our results and follow with a discussion in section \ref{sec:discussion}.

\section{The velocity correlation function} \label{sec:theory}

\subsection{Linear theory}

For the velocity field $\vecv$, we define the velocity correlation tensor $\Psi_{ij}(\vecr) = \langle v_i(\vecx)v_j(\vecx+\vecr) \rangle$. As the velocity field can be well approximated as curl-free, knowledge of the velocity divergence field $\theta(\vecx)$ is sufficient to describe the velocity field;
\begin{equation}
\theta(\vecx) \equiv  -\frac{\nabla \boldsymbol{\cdot} \vecv}{aHf}.
\end{equation}
The Fourier transform (denoted by tildes) of $\theta(\vecx)$ is
\begin{equation}\label{eq:theta}
 \tilde{\theta}(\veck) = \rmi\veck \boldsymbol{\cdot} \frac{\tilde\vecv(\veck)}{aHf},
 \end{equation}
and using linear theory, $\tilde{\theta}(\veck)=\tilde{\delta}_{\rm L}(\veck)$, with $\delta_{\rm L}$ being the linear matter density fluctuation. Then,

\begin{eqnarray}\label{eq:vv}
   \langle v_i(\vecr_1) v_j(\vecr_2) \rangle
&=& \int \frac{{\rm d}^3k}{(2\pi)^3}e^{-\rmi\veck \boldsymbol{\cdot} \vecr_1}  \int \frac{{\rm d}^3k'}{(2\pi)^3}e^{-\rmi\veck' \boldsymbol{\cdot} \vecr_2} \langle  \tilde{v}_i(\veck) \tilde{v}_j(\veck') \rangle \\
&=& -(aHf)^2  \int \frac{{\rm d}k}{2\pi^2}P_{\rm L}(k) \left[(\delta^{\rm D}_{ij}-\hat{r}_i\hat{r}_j)\frac{j_0'(kr)}{kr} + \hat{r}_i\hat{r}_j j_0''(kr)\right]\\
&=& \hat{r}_i\hat{r}_j \Psi_{\parallel}^{vv} + (\delta^{\rm D}_{ij}-\hat{r}_i\hat{r}_j)\Psi_{\perp}^{vv} .
\end{eqnarray}

\noindent Here, $\vecr=\vecr_1 - \vecr_2$, $\hat{r}_i$ is the $i$-th component of the unit vector $\hat{\vecr}=\vecr/r$, $j_n(kr)$ is the spherical Bessel function of $n$-th order, $P_{\rm L}(k)$ is the linear matter power spectrum, $\langle \tilde{\delta}_{\rm L}(\veck) \tilde{\delta}_{\rm L}(\veck') \rangle = (2\pi)^3 \delta^{\rm D} (\veck+\veck')P_{\rm L}(k)$
and $\delta^{\rm D}$ is the Dirac delta function.
We have defined \citep{Gorski:1988,Dodelson:2003}
\begin{eqnarray}
\label{eq:Psi_para} \Psi_{\parallel}^{vv}  &=&  -(aHf)^2  \int \frac{{\rm d}k}{2\pi^2 }j''_0(kr) P_{\rm L}(k) = (aHf)^2\int \frac{{\rm d}k}{2\pi^2 }\left[j_0(kr)-2\frac{j_1(kr)}{kr} \right]P_{\rm L}(k)\\
\label{eq:Psi_perp} \Psi_{\perp}^{vv} &=& -(aHf)^2  \int \frac{{\rm d}k}{2\pi^2}\frac{j_0'(kr)}{kr}P_{\rm L}(k) = (aHf)^2  \int \frac{{\rm d}k}{2\pi^2}\frac{j_1(kr)}{kr}P_{\rm L}(k).
\end{eqnarray}
See Appendix \ref{appendix:derivation} for the complete derivation.
In linear theory, $\langle \tilde{\theta}(\veck_1) \tilde{\theta}(\veck_2) \rangle = (2\pi)^3 \delta^{\rm D}(\veck_1+\veck_2)P_{\rm L}(k)$.
We can only observe the line-of-sight velocity component $u\equiv \vecv\boldsymbol{\cdot}\hat{\boldsymbol{n}}$, where $\hat{\boldsymbol{n}}$ is the unit vector pointing along the line-of-sight to the observed point (in this section, $\hat{\boldsymbol{n}}=\hat{\vecr}_1$ or $\hat{\boldsymbol{n}}=\hat{\vecr}_2$).
The two-point function of $u$ is obtained by contracting $\langle v_i(\vecr_1) v_j(\vecr_2) \rangle$ with $\hat{r}_{1,i}$ and $\hat{r}_{2,j}$ \citep{Dodelson:2003}:
\begin{eqnarray}
 \nonumber   \langle u(\vecr_1) u(\vecr_2) \rangle &=& \hat{r}_{1,i} \langle v_i(\vecr_1) v_j(\vecr_2) \rangle \hat{r}_{2,j}\\
\label{eq:uu}   &=& [\Psi_{\perp}^{vv}\cos{\theta_{12}}+(\Psi_{\parallel}^{vv}-\Psi_{\perp}^{vv})\cos{\theta_1}\cos{\theta_2}],
\end{eqnarray}
where we define the angles $\cos{\theta_1}\equiv \hat\vecr_1 \boldsymbol{\cdot} \hat\vecr$, $\cos{\theta_2}\equiv{\hat\vecr_2 \boldsymbol{\cdot} \hat\vecr}$, and $\cos{\theta_{12}}={\hat\vecr_1 \boldsymbol{\cdot} \hat\vecr_2}$.

\cite{Gorski:1989} developed a set of statistics $\psi_1$ and $\psi_2$ which can be obtained from galaxy PV surveys,
\begin{eqnarray}
 \label{eq:psi1}   \psi_1(r) &\equiv& \frac{\sum_{\rm D} w_1 w_2 u_1 u_2 \cos{\theta_{12}}}{\sum_{\rm D} w_1 w_2 \cos^2{\theta_{12}}} \\
 \label{eq:psi2}   \psi_2(r) &\equiv& \frac{\sum_{\rm D} w_1 w_2 u_1 u_2 \cos{\theta_1}\cos{\theta_2}}{\sum_{\rm D} w_1 w_2 \cos{\theta_{12}}\cos{\theta_1}\cos{\theta_2}},
\end{eqnarray}
where the sum $\sum_{\rm D}$ is taken over galaxy pairs in the data sample with separation $r$, and $w_i$ is the weight factor of each galaxy; typically the inverse of the selection function. 
We approximate the ensemble expectation values as:
\begin{eqnarray}
\label{eq:p1ea}\langle \psi_1(r) \rangle &\simeq&
\frac{\int {\rm d}^3r_1{\rm d}^3r_2 \delta^{\rm D}(|\vecr_2-\vecr_1|-r)w(\vecr_1)w(\vecr_2)\langle[1+\delta^{\rm g}(\vecr_1)][1+\delta^{\rm g}(\vecr_2)]u_1(\vecr_1)u_2(\vecr_2)\rangle\cos{\theta_{12}}}
{\int {\rm d}^3r_1{\rm d}^3r_2 \delta^{\rm D}(|\vecr_2-\vecr_1|-r)w(\vecr_1)w(\vecr_2)\langle[1+\delta^{\rm g}(\vecr_1)][1+\delta^{\rm g}(\vecr_2)]\rangle\cos^2{\theta_{12}}} ,\\
\nonumber & & \\
\label{eq:p2ea} \langle \psi_2(r) \rangle &\simeq&
\frac{\int {\rm d}^3r_1{\rm d}^3r_2 \delta^{\rm D}(|\vecr_2-\vecr_1|-r)w(\vecr_1)w(\vecr_2)\langle[1+\delta^{\rm g}(\vecr_1)][1+\delta^{\rm g}(\vecr_2)]u_1(\vecr_1)u_2(\vecr_2)\rangle\cos{\theta_{1}}\cos{\theta_{2}}}
{\int {\rm d}^3r_1{\rm d}^3r_2 \delta^{\rm D}(|\vecr_2-\vecr_1|-r)w(\vecr_1)w(\vecr_2)\langle[1+\delta^{\rm g}(\vecr_1)][1+\delta^{\rm g}(\vecr_2)]\rangle\cos{\theta_{12}}\cos{\theta_1}\cos{\theta_2}} .
\end{eqnarray}

\noindent We note that the operation is not exact because we have taken the ensemble average of the numerator and denominator separately, but in general $\langle X/Y \rangle \neq \langle X \rangle / \langle Y \rangle$. However, using $240$ mock sample realizations (described in Section \ref{sec:floats}), we have verified that the equality holds within $0.1$ per cent for the specific cosmological fields considered in this work. Thus, we apply the ensemble average separately to the numerator and denominator. See Appendix \ref{appendix:supp} for more details on this point.

We can see that the density term $\delta^{\rm g}(\vecr)$, the galaxy overdensity, explicitly appears. This is because $\psi_{1,2}$ are defined as pair-weighted quantities. In the original paper,
they assumed $\delta^{\rm g} \ll 1$
to eliminate the contribution of $\delta^{\rm g}(\vecr_{1}), \delta^{\rm g}(\vecr_{2})$ to $\psi_{1,2}$ \citep{Gorski:1989}. In this case, using $\langle u(\vecr_1)u(\vecr_2)\rangle$ from equation (\ref{eq:uu}), we obtain

\begin{eqnarray}
\label{eq:lp1} \langle \psi_1(r) \rangle &=&  A(r) \Psi_{\parallel}^{vv}+ [1-A(r)]\Psi_{\perp}^{vv}\\
\label{eq:lp2} \langle \psi_2(r) \rangle &=&  B(r) \Psi_{\parallel}^{vv}+ [1-B(r)]\Psi_{\perp}^{vv},
\end{eqnarray}
where we have defined
\begin{eqnarray}
\label{eq:A}    A(r) &=& \frac{\sum_{\rm D} w_1 w_2 \cos{\theta_1}\cos{\theta_2}\cos{\theta_{12}}}{\sum_{\rm D} w_1 w_2 \cos^2{\theta_{12}}}\\
\label{eq:B}    B(r) &=& \frac{\sum_{\rm D} w_1 w_2 \cos^2{\theta_1}\cos^2{\theta_2}}{\sum_{\rm D} w_1 w_2 \cos{\theta_1}\cos{\theta_2}\cos{\theta_{12}}}.
\end{eqnarray}
These formulae 
can be used to place constraints on $f$, by fitting the amplitude of $\psi_{1}$, $\psi_{2}$. However, the additional term, which is the density-velocity correlation, may have an impact, especially when the galaxy bias $b$ is large and when smaller scales -- where $\delta$ is larger -- are used \citep{Sheth:2008,Okumura:2014}.

The exact expansion of the density weighted velocity in equations (\ref{eq:p1ea}) and (\ref{eq:p2ea}) is,
\begin{equation}
\label{eq:dvv}
\langle [1+\delta^{\rm g}(\vecr_1)]u(\vecr_1)[1+\delta^{\rm g}(\vecr_2)]u(\vecr_2) \rangle =  \langle u(\vecr_1)u(\vecr_2) \rangle+  \langle \delta^{\rm g}(\vecr_1) u(\vecr_1) u(\vecr_2) \rangle+ \langle u(\vecr_1) \delta^{\rm g}(\vecr_2) u(\vecr_2) \rangle+ \langle \delta^{\rm g}(\vecr_1) u(\vecr_1) \delta^{\rm g}(\vecr_2) u(\vecr_2) \rangle .
\end{equation}
In the linear regime, both density and velocity field are Gaussian, thus the three-point averages are zero and the four-point average is decomposed into products of two-point averages. As a result, we have 
\begin{equation}
\label{eq:exp}     \langle [1+\delta^{\rm g}(\vecr_1)]u(\vecr_1)[1+\delta^{\rm g}(\vecr_2)]u(\vecr_2) \rangle
     =  [1+\langle \delta^{\rm g}(\vecr_1)\delta^{\rm g}(\vecr_2) \rangle] \langle u(\vecr_1)u(\vecr_2) \rangle
     + \langle \delta^{\rm g}(\vecr_1) u(\vecr_1) \rangle \langle  \delta^{\rm g}(\vecr_2) u(\vecr_2) \rangle
      + \langle \delta^{\rm g}(\vecr_1) u(\vecr_2) \rangle \langle  \delta^{\rm g}(\vecr_2) u(\vecr_1) \rangle .
\end{equation}

\noindent The first term on the right hand side is the quantity that has been previously considered. Note that the $[1+\langle \delta^{\rm g}(\vecr_1)\delta^{\rm g}(\vecr_2) \rangle]$ term cancels out in the expressions of $\psi_{1,2}$, due to the same term in the denominator.
The density-velocity term is calculated as

\begin{eqnarray}
\langle \delta^{\rm g}(\vecr_1) v_i(\vecr_2) \rangle
&=& \int \frac{{\rm d}^3k}{(2\pi)^3}e^{-\rmi\veck\boldsymbol{\cdot}\vecr_1} \int \frac{{\rm d}^3k'}{(2\pi)^3}e^{-\rmi\veck'\boldsymbol{\cdot}\vecr_2} \langle \tilde{\delta}^{\rm g}(\veck) \tilde{v}_i(\veck')\rangle\\
&=& (aHfb) \int \frac{{\rm d}k}{2\pi^2}P_{\rm L}(k) \frac{\rm d}{{\rm d}r_i}j_0(kr)\\
&=& b\hat{r}_i \Psi^{\delta v},
\end{eqnarray}
where we define 
\begin{equation}
   \Psi^{\delta v}(r) = -\frac{aHf}{2\pi^2}\int {\rm d}k\,kP(k)j_1(kr).
\end{equation}
Finally we can write 
\begin{equation}
    \langle \delta^{\rm g}(\vecr_1) u(\vecr_2) \rangle = \langle \delta^{\rm g}(\vecr_1) v_i(\vecr_2) \rangle \tilde{r}_{2,i}
    = b\Psi^{\delta v} \cos{\theta_2}
\end{equation}
Similarly, we find
\begin{equation}
    \langle u(\vecr_1) \delta^{\rm g}(\vecr_2) \rangle = \langle \delta^{\rm g}(\vecr_2) v_i(\vecr_1) \rangle \tilde{r}_{1,i}
    = - b\Psi^{\delta v} \cos{\theta_1}
\end{equation}
by noticing that $-\rmi\veck \boldsymbol{\cdot} \vecr$ in the above integration becomes $\rmi\veck\boldsymbol{\cdot}\vecr$ \citep{Adams:2017, Turner:2021}.

This is the same expression as equation (15) and (16) of \citet{Turner:2021} which is the galaxy-velocity correlation. 
The last term on the right hand side of equation (\ref{eq:exp}); $\langle \delta(\vecr_1) u(\vecr_1) \rangle \langle \delta(\vecr_2) u(\vecr_2) \rangle$, is zero because the fields in the ensemble averages are evaluated at a common point and $j_1(0)=0$. Physically this is a consequence of there being no special direction of $\vecv(\vecr_1)$ at $\vecr_1$.
In summary, the density-weighted line-of-sight velocity correlation function reads
\begin{eqnarray}
 \langle [1+\delta^{\rm g}(\vecr_1)]u(\vecr_1)[1+\delta^{\rm g}(\vecr_2)]u(\vecr_2) \rangle 
 &=& [1+\xi^{\rm g}(r)][\Psi_{\perp}^{vv}\cos{\theta_{12}}+(\Psi_{\parallel}^{vv}-\Psi_{\perp}^{vv})\cos{\theta_1}\cos{\theta_2}] - b^2(\Psi^{\delta v})^2 \cos{\theta_1}\cos{\theta_2}\\
 &=&  [1+\xi^{\rm g}(r)]\left[\Psi_{\perp}^{vv}\cos{\theta_{12}}+\left(\left[\Psi_{\parallel}^{vv}-\frac{b^2}{1+\xi^{\rm g}(r)}(\Psi^{\delta v})^2\right]-\Psi_{\perp}^{vv}\right)\cos{\theta_1}\cos{\theta_2}\right],
\end{eqnarray}
where $\xi^{\rm g}(r)$ is the galaxy two-point correlation function.

\noindent The quantity $\Psi^{\delta v}$ is understood to act to reduce $\Psi_{\parallel}^{vv}$ as
\begin{equation} \Psi_{\parallel}^{vv}(r) \mapsto \Psi_{\parallel}^{vv}(r) - \frac{b^2}{1+\xi^{\rm g}(r)}(\Psi^{\delta v}(r))^2 \, \end{equation}
whereas $\Psi_{\perp}^{vv}$ is unaffected. This was noted in \citet{Sheth:2008}. The density-velocity cross correlation therefore solely contributes to the parallel component, which physically corresponds to infall motion in the linear regime. The component perpendicular to the separation vector; $\langle \delta_1 v_{2\perp} \rangle$, is zero due to parity. 
Finally, the $\psi_{1,2}$ statistics are modified as
\begin{eqnarray}
\label{eq:lin_p1} \langle \psi_1(r) \rangle &=&   A(r) \left(\Psi_{\parallel}^{vv}-\frac{b^2(\Psi^{\delta v})^2}{1+\xi^{\rm g}(r)}\right) + [1-A(r)]\Psi_{\perp}^{vv}\\
\label{eq:lin_p2} \langle \psi_2(r) \rangle &=&   B(r) \left(\Psi_{\parallel}^{vv}-\frac{b^2(\Psi^{\delta v})^2}{1+\xi^{\rm g}(r)}\right) + [1-B(r)]\Psi_{\perp}^{vv}.
\end{eqnarray}
The contribution from $\Psi^{\delta v}$ is negative because the galaxy velocities are directed towards overdensities. On the other hand, $\langle vv \rangle$ is positive, because it is a volume-weighted quantity. Two infinitesimal volumes separated by $\vecr$ tend to move in similar directions, as both are on the same velocity flow sourced by the underlying matter overdensity field. In linear theory, $\Psi_{\parallel}^{vv}$ is proportional to $\delta_{\rm L}^2$. The additional term $(\Psi^{\delta v})^2$ goes as $\delta_{\rm L}^4$ in the numerator, so on small scales where the matter density fluctuation is larger, its contribution becomes important. Although the significance of this term is somewhat canceled by the factor of $1 + \xi^{g}$ in the denominator, it is important to determine its significance. This will be the focus of the following sections. 

In this paper we adopt the plane-parallel limit, in which case $\cos{\theta_1}=\cos{\theta_2}=\mu$ and $\cos{\theta_{12}}=1$,
greatly simplifying $\psi_1$ and $\psi_2$. In this limit we can write $A(r)\sim 1/3$ and $B(r) \sim 3/5$,
$\langle \psi_1(r)\rangle = \frac{1}{3} \Psi_{\parallel}^{vv}(r)+\frac{2}{3}\Psi_{\perp}^{vv}(r)$ and $\langle \psi_2(r)\rangle = \frac{3}{5} \Psi_{\parallel}^{vv}(r)+\frac{2}{5}\Psi_{\perp}^{vv}(r)$. The former is the well-known 1-D velocity correlation function. 
Existing PV survey data is limited to the local universe, up to $z \simeq 0.1$, or comoving distances $d_{\rm c} \sim 300 h^{-1} \, {\rm Mpc}$. The maximum scale used in correlation function analyses is of order $\sim 100 h^{-1} \, {\rm Mpc}$ and may be affected by the breakdown of the plane-parallel limit. However, the main scope of this paper is to investigate non-linear effects that are relevant on relatively small scales. As we will see, the statistical constraining power at $\sim 100 \, h^{-1} \, {\rm Mpc}$ is extremely weak, and thus the wide-angle effect does not affect our arguments. This is likely why the plane-parallel approximation is so effective in many applications to data despite being divorced from reality; the smallest scale modes provide the majority of the constraining power. Also note that in real-space analysis, the pair $\psi_1$ and $\psi_2$ can be transformed into $\Psi_{\parallel}^{vv}$ and $\Psi_{\perp}^{vv}$, which are readily estimated from the theory. These quantities are practically free from the complexities of the plane-parallel approximation and also the survey geometry, since this information is incorporated into the $A(r)$, $B(r)$ coefficients. 
Expanding the velocity power spectrum in redshift space, accounting for the wide angle effect, is rather complicated because of additional angular dependence \citep{Dam:2021}. For future data like DESI \citep{Saulder:2023}, which will probe larger volumes, the plane-parallel approximation may not be sufficiently accurate.

\subsection{Non-linearity}
\label{sec:nonlin}

In the previous subsection, linear theory was assumed. 
However, below a certain scale, non-linear effects become important. We now present the higher order corrections based on standard perturbation theory.
In this language, the overdensity $\tilde{\delta}(\veck)$ and velocity divergence $\tilde{\theta}(\veck)$ are perturbatively expanded as \citep{Makino:1992,Jain:1994,Bernardeau:2002}
\begin{eqnarray}
\delta &=&  \delta^{(1)} + \delta^{(2)} +  \delta^{(3)} + \ldots\\
\theta &=&  \theta^{(1)} + \theta^{(2)} + \theta^{(3)} +  \ldots
\end{eqnarray}
The first terms are the linearized quantity, $\delta^{(1)} = \theta^{(1)} = \delta_{\rm L}$.
The $n$-th term are order of $\delta_{\rm L}^n$ and can be expressed as the convolution of the lower orders with the coupling kernels,
\begin{eqnarray}
\tilde{\delta}^{(n)}(\veck)&=&\frac{1}{n!}\int \frac{{\rm d}^3k_1}{(2\pi)^3}\cdots \frac{{\rm d}^3k_n}{(2\pi)^3}(2\pi)^3\delta^{\rm D}(\veck_1+\cdots + \veck_n -\veck)F^{(n)}(\veck_1,\cdots,\veck_n)\delta_{\rm L}(\veck_1)\cdots\delta_{\rm L}(\veck_n) \\
\tilde{\theta}^{(n)}(\veck)&=&\frac{1}{n!}\int \frac{{\rm d}^3k_1}{(2\pi)^3}\cdots \frac{{\rm d}^3k_n}{(2\pi)^3}(2\pi)^3\delta^{\rm D}(\veck_1+\cdots + \veck_n -\veck)G^{(n)}(\veck_1,\cdots,\veck_n)\delta_{\rm L}(\veck_1)\cdots\delta_{\rm L}(\veck_n).
\end{eqnarray}
The kernels are obtained solving the continuity equation, Euler equation, and Poisson equation. The explicit forms are given by \citet{Bernardeau:2002}.
Using these, the auto-power spectrum of the velocity divergence field becomes
\begin{equation}
    P_{\theta\theta}(k) = P_{\rm L}(k) 
+ 2 \int_{\vecq}P_{\rm L}(q)P_{\rm L}(|\veck-\vecq|)\left[G^{(2)}(\vecq,\veck-\vecq)\right]^2. + 6P_{\rm L}(k)\int_{\vecq} P_{\rm L}(q)\left[G^{(3)}(\veck,\vecq,-\vecq)\right]
\end{equation}
The integral notation $\int_{\vecq}$ means $(2\pi)^{-3}\int{{\rm d}^3q}$.
The integrations can be performed analytically, yielding
\begin{equation}\label{eq:Pthetatheta}
    P_{\theta\theta}(k) = P_{\rm L}(k) + \frac{k^3}{2\pi^2}\left[P_{\theta\theta}^{(22)}(k) + P_{\theta\theta}^{(13)}(k)\right],
\end{equation}
where $P_{\theta\theta}^{(22)}$ and $P_{\theta\theta}^{(13)}$ are the next-to-leading order contributions,
\begin{eqnarray}
  P_{\theta\theta}^{(22)}(k) &=& \int q'^2 {\rm d}q' P_{\rm L}(kq') \int^{1}_{-1} {\rm d}\mu' \frac{1}{2}I_{\theta\theta}^{(22)}(q',\mu')P_{\rm L}(ky) \label{eq:P_22}\\
  P_{\theta\theta}^{(13)}(k) &=& P_{\rm L}(k) \int q'^2{\rm d}q' J_{\theta\theta}^{(13)}(q')P_{\rm L}(kq')\label{eq:P_13},
\end{eqnarray}
with $q'=q/k$, $\mu'=(\veck \cdot \vecq)/(kq)$ and $y=\sqrt{1+q'^2-2q'\mu'}$ and the kernels $I_{\theta\theta}^{(22)}(q',\mu')$ and $J_{\theta\theta}^{(13)}(q')$ are given in \citet{Vlah:2012}; specifically, $I_{\theta\theta}^{(22)}(q',\mu')$ is $2f_{11}$ in their paper and  $J_{\theta\theta}^{(13)}(q)$ is $6g_{11}(q)/q^2$. Both of these terms scale as $P_{\rm L}^2 \propto \delta_{\rm L}^4$.

The cross power spectrum between the density and velocity divergence becomes
\begin{equation}
P_{\delta\theta}(k) = P_{\rm L}(k) 
+2\int_{\vecq}P_{\rm L}(q)P_{\rm L}(|\veck-\vecq|)\left[F^{(2)}(\vecq,\veck-\vecq)G^{(2)}(\vecq,\veck-\vecq)\right] + 3P_{\rm L}(k)\int_{\vecq} P_{\rm L}(q)\left[G^{(3)}(\veck,\vecq,-\vecq) +F^{(3)}(\veck,\vecq,-\vecq)\right].
\end{equation}
It can be expressed as
\begin{equation}
P_{\delta\theta}(k) = P_{\rm L}(k)+\frac{k^3}{2\pi^2}\left[P_{\delta\theta}^{(22)}(k) + P_{\delta\theta}^{(13)}(k)\right],
\end{equation}
where
\begin{eqnarray}
  P_{\delta\theta}^{(22)}(k) &=& \int q'^2{\rm d}q' P_{\rm L}(kq') \int^{1}_{-1} {\rm d}\mu' \frac{1}{2}I_{\delta\theta}^{(22)}(q',\mu')P_{\rm L}(ky)\\
  P_{\delta\theta}^{(13)}(k) &=& P_{\rm L}(k) \int q'^2{\rm d}q' J_{\delta\theta}^{(13)}(q')P_{\rm L}(kq').
\end{eqnarray}

\noindent Without the linear assumption, the cubic contributions to equation (\ref{eq:dvv}) are not trivial.
The relevant terms are $\langle \delta^{\rm g}(\vecr_1)\theta(\vecr_1)\theta(\vecr_2)\rangle$ 
and $\langle \theta(\vecr_1)\delta^{\rm g}(\vecr_2)\theta(\vecr_2)\rangle$, which can be expanded as 
$\langle \delta \theta \theta \rangle = \langle \delta^{(2)}\theta^{(1)}\theta^{(2)}\rangle+\langle \delta^{(1)}\theta^{(2)}\theta^{(1)}\rangle+\langle \delta^{(1)}\theta^{(1)}\theta^{(2)}\rangle$. Carrying out the Fourier transform of these terms, we obtain

\begin{eqnarray}\label{eq:dvv}
   \langle \delta^{\rm g}(\vecr_1)v_i(\vecr_1) v_j(\vecr_2) \rangle
&=& \int \frac{{\rm d}^3k}{(2\pi)^3}e^{-\rmi\veck \boldsymbol{\cdot} \vecr_1}
\int \frac{{\rm d}^3k'}{(2\pi)^3}e^{-\rmi\veck' \boldsymbol{\cdot} \vecr_1}
\int \frac{{\rm d}^3k''}{(2\pi)^3}e^{-\rmi\veck'' \boldsymbol{\cdot} \vecr_2} \langle  \tilde{\delta}^{\rm g}(\veck)\tilde{v}_i(\veck') \tilde{v}_j(\veck'') \rangle \\
&=& \hat{r}_i\hat{r}_j b\Psi_{\parallel}^{\delta vv} + (\delta^{\rm D}_{ij}-\hat{r}_i\hat{r}_j) b \Psi_{\perp}^{\delta vv},
\end{eqnarray}
where
\begin{eqnarray}
\Psi_{\parallel}^{\delta vv}  &=&   (aHf)^2\int \frac{{\rm d}k}{2\pi^2 }\left[j_0(kr)-2\frac{j_1(kr)}{kr} \right]B_{\delta \theta \theta}(k)\\
\Psi_{\perp}^{\delta vv} &=&  (aHf)^2  \int \frac{{\rm d}k}{2\pi^2}\frac{j_1(kr)}{kr}B_{\delta \theta \theta}(k).
\end{eqnarray}
The bispectrum $B_{\delta \theta \theta}(k)$ up to $\delta_L^{4}$ is
\begin{equation}
    B_{\delta\theta\theta}(k) =  \frac{k^3}{2\pi^2}\left[B_{\delta\theta\theta}^{(112)}(k) + B_{\delta\theta\theta}^{(121)}(k)\right],
\end{equation}
where
\begin{eqnarray}
  B_{\delta\theta\theta}^{(112)}(k) &=& \int q'^2 {\rm d}q' P_{\rm L}(kq') \int^{1}_{-1} {\rm d}\mu' \frac{1}{2}I_{\delta\theta\theta}^{(112)}(q',\mu')P_{\rm L}(ky) \label{eq:B_112}\\
  B_{\delta\theta\theta}^{(121)}(k) &=& P_{\rm L}(k) \int q'^2{\rm d}q' J_{\delta\theta\theta}^{(121)}(q')P_{\rm L}(kq')\label{eq:B_121},
\end{eqnarray}
The integration kernels can be found  in \citet{Vlah:2012}; specifically, $I_{\delta\theta\theta}^{(112)}(q',\mu')$ is $2f_{22}$ in their paper and   $J_{\delta\theta\theta}^{(121)}(q)$ is $6g_{10}(q)/q^2$.

One finds that $\langle \delta(\vecr_1)\theta(\vecr_1)\theta(\vecr_2)\rangle$ = $\langle \theta(\vecr_1)\delta(\vecr_2)\theta(\vecr_2)\rangle$ and that
the angular dependencies of $\langle vv \rangle$ and $\langle dvv \rangle$ (equations (\ref{eq:vv}) and (\ref{eq:dvv})) are identical.
As a consequence, the full expression for the non-linear $\psi_1$ and $\psi_2$ becomes
\begin{eqnarray}
\langle \psi_1(r) \rangle &=&   A(r) \left(\Psi_{\parallel}^{vv} + 2b\Psi_{\parallel}^{\delta vv}-\frac{b^2(\Psi^{\delta v})^2}{1+\xi^{\rm g}(r)}\right) + [1-A(r)]\left(\Psi_{\perp}^{vv}+2b\Psi_{\perp}^{\delta vv} \right) \label{eq:psi1_nl}\\
\langle \psi_2(r) \rangle &=&   B(r) \left(\Psi_{\parallel}^{vv} + 2b\Psi_{\parallel}^{\delta vv}-\frac{b^2(\Psi^{\delta v})^2}{1+\xi^{\rm g}(r)}\right) + [1-B(r)]\left(\Psi_{\perp}^{vv}+2b\Psi_{\perp}^{\delta vv}\right) \label{eq:psi2_nl}.
\end{eqnarray}

The modifications from the linear version are; (1) $\Psi_{\parallel}^{vv}$ and $\Psi_{\perp}^{vv}$ are calculated with PT. Specifically, $P_L(k)$ in equations (\ref{eq:Psi_para}) and (\ref{eq:Psi_perp}) is replaced with $P_{\theta\theta}(k)$ of equation (\ref{eq:Pthetatheta}). (2) additional terms $\Psi_{\parallel}^{\delta vv}$ and $\Psi_{\perp}^{\delta vv}$ appear.
In real space, all terms are proportional to $f\sigma_8$. We will use $N$-body simulations to measure the correlation functions and fit them with the theoretical template, varying $f\sigma_8$ to check how these additional terms affect the best-fitting value. In linear theory, the pair $(\psi_1,\psi_2)$ and $(\Psi_{\parallel}^{vv},\Psi_{\perp}^{vv})$ contain the same cosmological information, as they are related via the functions $A(r)$ and $B(r)$ which are purely survey-specific quantities. However, the conversion becomes more complicated in the non-linear treatment due to the additional terms. Therefore, in this paper, we use the direct observable $(\psi_1,\psi_2)$ for the fitting, as is common in the literature.

Equations (\ref{eq:psi1_nl}) and (\ref{eq:psi2_nl}) now contain all effects up to $\delta_{\rm L}^4$ for $\psi_1$ and $\psi_2$. These quantities do not require the plane-parallel limit and can be used for any survey geometry. In Fourier space analysis, where the plane-parallel approximation is used, the line-of-sight momentum power spectrum is formulated by \citet{Vlah:2012} and the explicit form up to $\delta_{L}^4$ can be found in Appendix B of \citet{Howlett:2019bky}. While their expression is in redshift-space, the real-space part of the density-weighted velocity power spectrum is equivalent to $P_{11}$ as defined in those works. There are seven terms, one being the linear $\langle vv \rangle$ and the other six corresponding to non-linear contributions. Of these, $I_{11}$ and $J_{11}$ express the non-linear effects of $\langle vv \rangle$ corresponding to equations (\ref{eq:P_22}) and (\ref{eq:P_13}), $I_{22}$ and $J_{10}$ are the three-point contributions corresponding to $\Psi^{\delta vv}$,
and $I_{13}$ and $I_{31}$ are the density-velocity contributions corresponding to $\Psi^{\delta v}$.

\begin{figure}
\centering 
\includegraphics[width=0.45\textwidth]{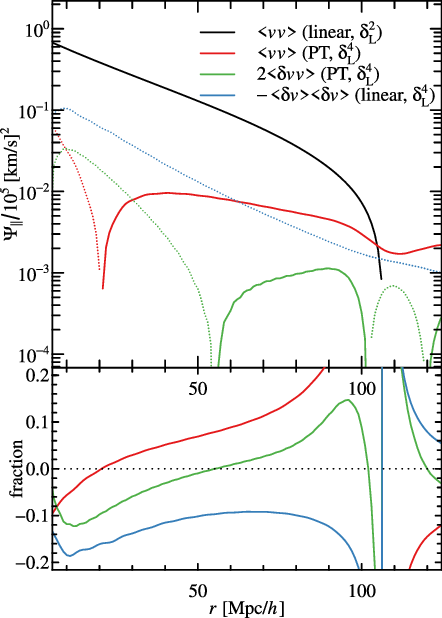}
\includegraphics[width=0.45\textwidth]{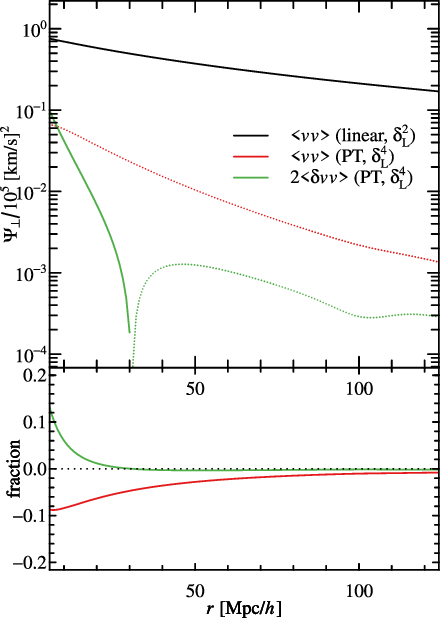}\\
\includegraphics[width=0.45\textwidth]{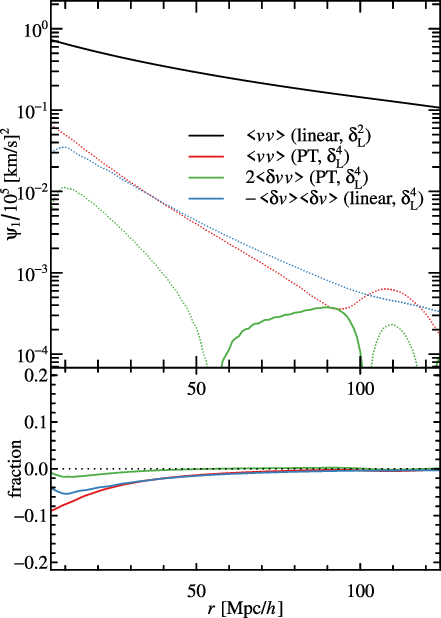}
\includegraphics[width=0.45\textwidth]{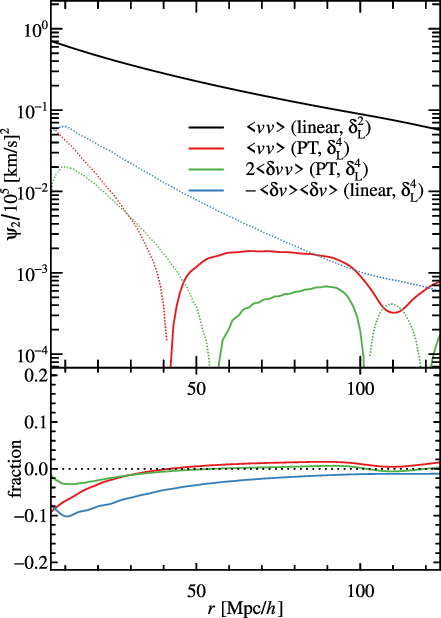}
\caption{\label{figure:theory}The velocity correlation functions, $\Psi_{\parallel}$, $\Psi_{\perp}$ (top panels), and $\psi_1$, $\psi_2$ (bottom panels). The solid line is the prediction from the linear theory $P_{\theta\theta}=P_{\rm L}$. The red lines are the contributions from $P_{\theta\theta}$ to $\Psi_{\parallel,\perp}^{vv}$ other than the linear part. The blue lines are contributions from the density-velocity correlation $\Psi^{\delta v}$. Note that it only modifies $\Psi_{\parallel}^{vv}$ and $\psi_{1,2}$ (which is a linear sum of $\Psi_{\parallel}$ and $\Psi_{\perp}$). The green lines are contributions from the three-point correlations $\Psi_{\parallel,\perp}^{\delta vv}$. On the bottom half of each panel, the relative contribution to the linear theory is presented. As the plot is logarithmic, negative values are expressed as dotted segments.}
\end{figure}

In Figure \ref{figure:theory} we present $\Psi_{\parallel}^{vv}$ and $\Psi_{\perp}^{vv}$ as a function of separation $r$. The black lines indicate the linear theory prediction and red the correction from the non-linear terms ($P_{\theta\theta}^{(22)}$ and $P_{\theta\theta}^{(13)}$). 
The sum of these higher order terms is negative\footnote{This is caused by the specific functional forms of $I_{\theta\theta}^{(22)}$ and $J_{\theta\theta}^{(13)}$. See figure 2 of \citet{Dam:2021}}, thus we expect them to contribute negatively to the overall prediction, as seen in the top right panel (red line), because $j_1(kr)$ is almost always positive. On the other hand, in the top left panel, their contribution to $\Psi_{\parallel}^{vv}$ is positive for scales larger than $20 \, h^{-1} \, {\rm Mpc}$.
This is due to the oscillatory nature of the spherical Bessel function combination $j_0(kr)-2j_1(kr)/(kr)$ in equation (\ref{eq:Psi_perp}). It is a diminishing trigonometric function and the period depends on the scale $r$. As a result, the tendency to be positive or negative is different for large and small $r$, and this causes a switch of the sign of the resulting integration.
Therefore, the effect of $P_{\theta\theta}^{(22)}$ and $P_{\theta\theta}^{(13)}$ on $f\sigma_8$ constraints will differ for $\Psi_{\parallel}^{vv}$ and $\Psi_{\perp}^{vv}$; using only $\Psi_{\parallel}^{vv}$ ($\Psi_{\perp}^{vv}$) and taking the linear theory prediction will result in an {\it over-estimation} ({\it under-estimation}) of $f\sigma_8$, implying the need to use both components to infer $f\sigma_8$.
If the PT-corrected $P_{\theta\theta}$ is used, the former will give a lower $f\sigma_8$ compared to linear theory, and the latter will give a higher $f\sigma_8$. 
Although these two contributions act to cancel the non-linear effect, the latter has a stronger $S/N$, meaning that the resulting $f\sigma_8$ is systematically biased higher. 

In the bottom panels of Figure \ref{figure:theory}, the statistic $\psi_1$ and $\psi_2$ are presented, where the plane-parallel approximation has been used. The effect of $\Psi^{\delta v}$ (blue lines) and $\Psi^{\delta vv}$ (green lines) enters both of $\psi_1$ and $\psi_2$ as negative. The effect of the perturbative $P_{\theta\theta}$ (red line) is mostly negative, but it is positive on $\leq 50 \ {\rm Mpc}/h$ for $\psi_2$, which is attributed to a larger contribution from $\Psi_{\parallel}$.
Overall, all the effects are approximately $3$--$10$ per cent over the range $20 \leq r \leq 50 h^{-1} \, {\rm Mpc}$, roughly consistent with the expected value of $\delta_{\rm L}^4$ compared to the linear theory $\delta_{\rm L}^2$ on these scales.
We have also calculated the next-to-leading order PT for $\delta v$. However, it appears as quadrature, $(\delta_{\rm L}^2+\delta_{\rm L}^4)^2$, so contributes at lowest order $\delta_{\rm L}^6$ and we do not show the effect in Figure \ref{figure:theory}.

\section{Simulation Data} \label{sec:floats}
\subsection{Quijote simulation}
In this section, we measure the velocity correlation function using dark matter particles. As dark matter is unbiased, it is an ideal tracer to compare to the theoretical model in the previous section. For this purpose, we use the Quijote simulation suite \citep{Navarro:2020}. It is a set of $N$-body simulations spanning over various cosmological parameters. Each realisation has box size $(1 h^{-1} \, {\rm Gpc})^{3}$. As we are only interested in the capability of the linear and non-linear model to reproduce the input cosmology, we use fiducial parameters that correspond to the Planck $2018$ cosmology \citep{Aghanim:2018}: $(\Omega_m,\sigma_8,h,w)=(0.3175, 0.834, 0.6711, -1)$. We use $30$ snapshot boxes at $z=0$, and then split each box into eight sub-boxes, each being a $(500 \, h^{-1} \,  {\rm Mpc})^{3}$ sized cube, resulting in $N_{r} = 240$ realizations in total. This volume roughly corresponds to the volume of the planned DESI PV survey \citep{Saulder:2023}. First, we measure $\psi_1(r)$ and $\psi_2(r)$ using equations (\ref{eq:psi1}) and (\ref{eq:psi2}).
We assume the plane-parallel approximation ($\cos{\theta_{12}}=1$, $\cos{\theta_1}=\mu$, and $u_1$ is the particle velocity along $z$-axis). 
We use $N_{\rm bin}=15$ bins for 
 the $3$-D separation over $[0,90] h^{-1} \, {\rm Mpc}$ with $\Delta r = 6 h^{-1} \, {\rm Mpc}$. The weight of each particle is fixed as $w = 1$. For the purpose of parameter fitting in Section \ref{sec:fitting}, we construct a  covariance matrix from the $N_{r} = 240$ realisations. The data vector $\boldsymbol{d}=[\boldsymbol{\psi}_1,\boldsymbol{\psi}_2]$ is a $2\times N_{\rm bin}$ component vector, where $\boldsymbol{\psi}_{1,2}$ are the $N_{\rm bin}$-element vector of the measurements $\psi_{1,2}(r)$. The covariance matrix for $\boldsymbol{d}$ is calculated as 
\begin{equation}
 C_{ij}= \frac{1}{N_{r}-1} \sum_{k=1}^{N_{r}}{(d_i - \bar{d}_i)(d_j - \bar{d}_j)} ,
\end{equation}
\noindent where $\bar{d}_{i}$ is the mean value of the $i^{\rm th}$ component of the vector $\boldsymbol
{d}$ from the realisations. Due to very strong correlations between different bins, the covariance matrix is dominated by the off-diagonal component and the inverse operation is found to be unstable. We therefore reduce the off-diagonal elements by $0.5$ per cent to make the covariance matrix stable to inverse operation. We follow \citet{Hartlap:2006kj} to correct the covariance matrix for the inaccuracy due to the limited number of realisations.

\subsection{Horizon Run 4}

Actual PV surveys utilize galaxies as tracers.
The second mock dataset that we use is Horizon Run 4 \citep{Kim:2015}, a cosmological scale dark matter simulation that evolved $6300^3$ particles in a $3150 \, h^{-1} \, {\rm Mpc}$ box from $z=100$ to $z=0$. The input cosmological parameters are $(\Omega_m,\sigma_8,h,w)=(0.26,1/1.26,0.72,-1)$, and the mass resolution is $3.0\times 10^9 M_{\odot}/h$. Haloes are identified in each snapshot box with the friends-of-friends algorithm with the linking length being $0.2$ of the mean particle separation, from which the merger tree is constructed. 
Galaxy properties are determined via the most bound particle (MBP)-galaxy abundance matching  approach \citep{Hong:2016}; the position and velocity of MBPs are assigned to the simulated galaxies. The minimum mass of the haloes is $2.7\times 10^{11} M_{\odot}/h$, which is equivalent to $30$ DM particles.
The simulation has a good mass resolution and occupies a large volume, hence captures the evolution of the large-scale modes. This feature in particular is desirable for studies of cosmic velocity, as this quantity is proportional to $\tilde{\delta}/k$. The additional factor of $k^{-1}$ increases the power on large scales relative to the density field.

We sort mock galaxies in descending order of mass, and select $38.6$ million galaxies from the heaviest 
in the $z=0$ snapshot box, corresponding to a number density comparable to the expected DESI PV survey. In the actual survey, the sampling density decreases as redshift increases, but we do not consider this effect.
To account for the velocity measurement error, we assume a Gaussian distributed uncertainty with width $\sigma=3000\ {\rm km}s^{-1}$. Random values drawn from the distribution are added to the velocity of each galaxy. This corresponds to a $15$ per cent fractional error of the distance measurement at $200 \, h^{-1} \, {\rm Mpc}$, which is comparable to the mean distance of the DESI PV survey \citep{Saulder:2023}.
A reasonable change of the value of the width $\sigma$ does not affect the measurement, because the random fluctuations tend to cancel out in the pair counting. 
In case of actual observations, the distance measurement error is proportional to distance, thus most constraining power will come from the nearby galaxies with smaller velocity errors. Also, for realistic data there will be an additional effect; the outer boundary occupies a larger surface area compared to the inner if the survey geometry is conical, leading to a net scatter of galaxies from outside the survey volume into the catalog unless the radial selection function decreases very rapidly. This will generate a bias for negative velocities with increasing redshift, i.e. the Malmquist bias. These effects must be accounted for prior to the extraction of the velocity correlation function from actual data. However, we do not adopt any velocity error gradient or cone effects in the mocks but only the typical velocity error and a fixed cubic volume of snapshot data, similarly to \citet{Turner:2021}. This is because we are only interested in how the non-linear terms alter the best-fitting value of $f\sigma_8$, and wish to separate out such additional effects, which are beyond the scope of this paper. 
We divide the $z=0$ snapshot box into $N_{r}=6^3=216$ sub-boxes, each being a $(525 h^{-1} \, {\rm Mpc})^{3}$ cube. We measure the data vector $\vecd$ and covariance matrix in the same manner as in the previous sub-section.

\section{Parameter Fitting}
\label{sec:fitting}

Next we fit the theoretical velocity correlation functions to the measured values extracted from the simulations. We measure $\psi_{1}(r)$, $\psi_{2}(r)$ from the mock data, then create theoretical templates for these statistics, generated from the cosmological parameters of Quijote and HR4 simulations. We then vary $f\sigma_8$, by multiplying the templates by factors of $[f\sigma_8/(f\sigma_8)_{\rm fid}]^2$. The fiducial values are $(f\sigma_8)_{\rm fid}=0.446$ and $0.381$ for Quijote and HR4, respectively.
We perform the fitting using two different theoretical templates: (1) linear theory $\langle \psi_{1} \rangle$, $\langle \psi_{2} \rangle$; equations (\ref{eq:Psi_para},\ref{eq:Psi_perp},\ref{eq:lp1},\ref{eq:lp2}) and (2) non-linear theory; equations  (\ref{eq:lin_p1},\ref{eq:lin_p2}) using the non-linear power spectra derived in Section \ref{sec:nonlin} and including the additional terms $b^2(\Psi^{\delta v})^2/(1+\xi^{\rm g}(r))$ and $\Psi^{\delta vv}$.
The additional term does not only depend on $f\sigma_8$; the numerator depends on $(f\sigma_8) (b\sigma_8)$ and so we also require knowledge of the bias parameter. It is common to add the information of density-density correlation function to break the degeneracy \citep{Adams:2017,Adams:2020dzw,Turner:2021}. However, as the purpose of this study is to investigate the bias of $f\sigma_8$ induced by neglecting the non-linear effects, we use the fixed values of $b\sigma_8$. For Quijote, $b\sigma_8=\sigma_8=0.834$, and for the specific HR4 galaxy sample considered in this work, $b\sigma_8=1.4\times0.794 = 1.111$, as inferred from the density two-point correlation function over 45 to 87 $h^{-1}$Mpc. For the HR4 mock galaxy tracers, higher order bias contributions are also present \citep{McDonald:2008,Saito:2014}. However, the higher order bias enters $b^2\langle \delta v \rangle ^2$ at ${\cal O}(\delta_{\rm L}^6)$, so we do not consider this effect here. \citet{Appleby:2023} found that the non-linear bias $b_2$ is consistent with zero for mock HR4 early-type galaxies (ETGs), selected to match a sub-set of ETGs in the SDSS catalog. The terms $A(r)$ and $B(r)$ are calculated with equations (\ref{eq:A},\ref{eq:B}) for Quijote and HR4 samples separately, but these are very close to $1/3$ and $3/5$ irrespective of scales.

To account for the suppression of the correlation functions caused by not capturing the large-scale modes due to the finite volume of the simulations, we set the lower limit of the integration, $k_{\rm min}$, in the Hankel transform in equations (\ref{eq:Psi_para}) and (\ref{eq:Psi_perp}) corresponding to the box size of each simulation. This will prevent the overestimation of the theoretical templates and consequent underestimation of $f\sigma_8$.
We note that the size of error bars may be still underestimated by missing the cosmic variance of wave modes larger than the simulation size, for which we do not correct in the present study. Analytic formulae for the covariance matrix has recently been studied by \citet{Blake:2024}.

Most observational and simulation studies of the velocity correlation functions set the minimum fitting scale between $20 \, h^{-1} \, {\rm Mpc} \leq r_{\rm min} \leq  60 \, h^{-1} \, {\rm Mpc}$. Due to the non-linear prescription, we expect that the model (2) reproduces the input $f\sigma_8$ better than model (1). It is common to set the minimum fitting scale to be a few tens of megaparsecs or below \citep{Wang:2018,Dupuy:2019,Turner:2021}.
Fitting increasingly smaller scales using the linear model inflicts a systematic bias to parameter reconstruction. Our aim is to examine at what minimum scale this bias becomes significant and problematic. To this end, we fix the maximum scale $r_{\rm max} = 90 \, h^{-1} \, {\rm Mpc}$ and vary the minimum scale. The choice of the maximum scale is not important because the constraining power is weakest at this point. The minimum scale is varied over the range $15 h^{-1} \, {\rm Mpc} \leq r_{\rm min} \leq 57 \, h^{-1} \,  {\rm Mpc}$, where we expect that the perturbative approach works. For each $r_{\rm min}$, we fit the velocity correlation functions $\psi_{1}(r)$, $\psi_{2}(r)$ over the range $[r_{\rm min}, r_{\rm max}]$. The data vector $\vecd$ and the associated covariance matrix are modified each time accordingly. We minimize the chi-squared function 
\begin{equation}
    \chi^2 = (\vecd-\vecd^{\rm th})^{\rm T} \mathbfss{C}^{-1} (\vecd-\vecd^{\rm th}) 
\end{equation}
to find the best-fitting $f\sigma_8$, and based on the likelihood ${\cal L} \propto \exp{\left[-\chi^2/2\right]}$, we find the $16$ and $84$ percentiles. For the theoretical model $\vecd^{\rm th}$, we use (1)  linear theory, (2) non-linear theory up to $\delta_{\rm L}^4$.

\begin{figure}
\centering 
\includegraphics[width=0.47\textwidth]{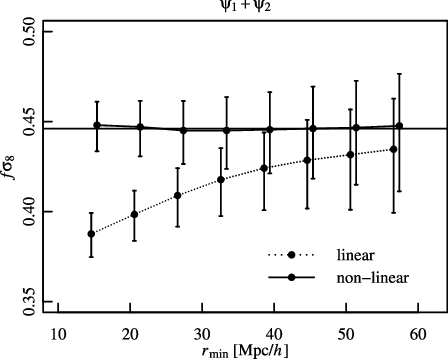}
\caption{The best-fitting value of $f\sigma_8$ obtained from the Quijote DM simulation. The fitting range is $[r_{\rm min},90] \, h^{-1} \, {\rm Mpc}$, and both $\psi_1(r)$ and $\psi_2(r)$ are fitted simultaneously. The dotted line assumes linear theory, whereas the solid line includes all $O(\delta_{\rm L}^4)$ terms.\label{figure:fsig8_as_rmin_Quijote} }
\end{figure}

\begin{figure}
\centering 
\includegraphics[width=0.47\textwidth]{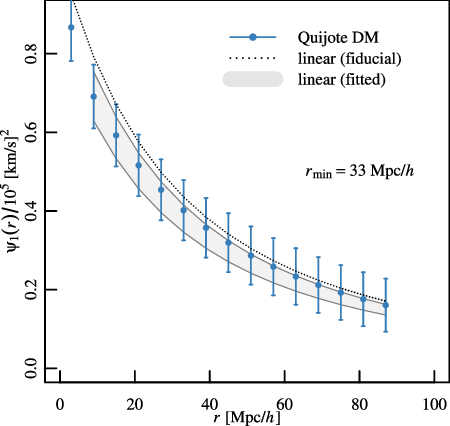}
\includegraphics[width=0.47\textwidth]{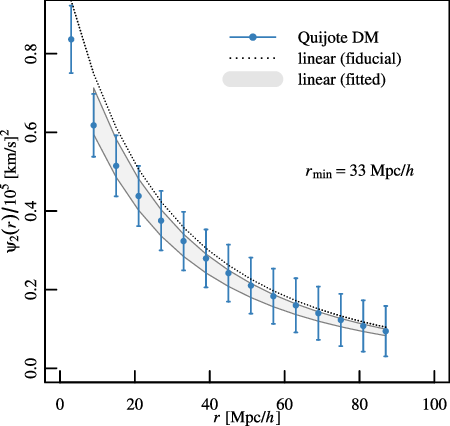}\\
\includegraphics[width=0.47\textwidth]{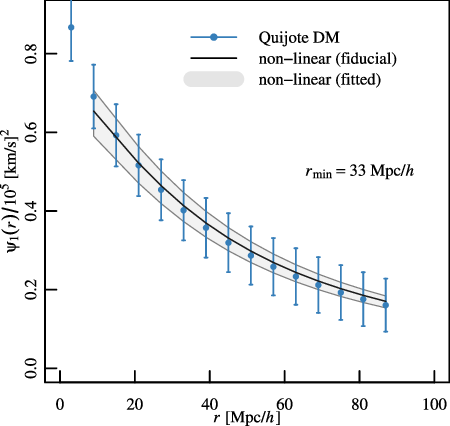}
\includegraphics[width=0.47\textwidth]{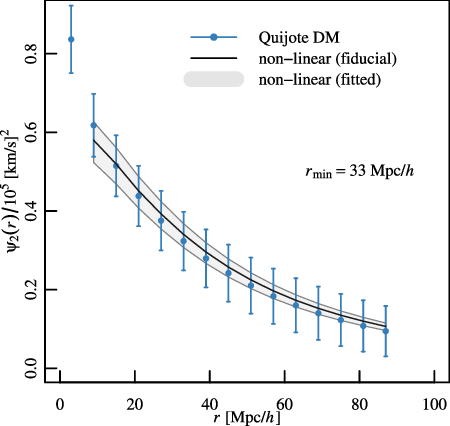}
\caption{The comparison between the measurements and the theoretical predictions. Top and bottom panels are for the linear theory and non-linear theory, respectively.  The measurements are presented as blue lines. The black lines are the fiducial theory ($f\sigma_8=0.446$). The gray regions indicate the resulting $1\sigma$ range of $f\sigma_8$, obtained from the simultaneous fitting to $\psi_1$ and $\psi_2$. \label{figure:bestfit_Quijote}}
\end{figure}

On large scales, both the linear and non-linear theories return the fiducial value. This can be seen in Figure \ref{figure:fsig8_as_rmin_Quijote}. For $r_{\rm min} \gtrsim 50 \, h^{-1} \, {\rm  Mpc}$, the two theories match the `true' value to within $1\sigma$. 
Figure \ref{figure:fsig8_as_rmin_Quijote} shows the best-fitting values of $f\sigma_8$ as a function of minimum scale for the Quijote simulation. On large scales, both templates result in $f\sigma_8$ consistent with the input value within $1\sigma$. Below $r_{\rm min} \sim 50 \, h^{-1} \, {\rm Mpc}$, using linear theory biases the reconstructed $f\sigma_8$ towards lower values, and the deviation systematically increases as $r_{\rm min}$ decreases. As explained  previously, this is because the linear theory overpredicts the velocity power spectrum, demanding that $f\sigma_8$ is smaller to compensate. For example, if we take the minimum scale as the mean separation of the first (third) bin; $r_{\min} = 21\ (33) \, h^{-1} \, {\rm Mpc}$, then the best-fitting value is $0.398\ (0.418)$ 
for the linear theory and $0.447 (0.445)$
for the non-linear theory compared to the fiducial value $(f\sigma_8)_{\rm Quijote}=0.446$. Figure \ref{figure:bestfit_Quijote} displays a comparison between the Quijote measurements and the best-fitting theoretical result, together with the models assuming the fiducial $(f\sigma_8)_{\rm Quijote}$. In the top panels, the fiducial linear theory is presented as dotted lines ($f\sigma_8=0.446$), and the grey shaded areas are the $1\sigma$ range obtained by fitting $f\sigma_{8}$ to the data ($0.398<f\sigma_8<0.435$).
The parameter estimated $f\sigma_{8}$ is significantly lower than the fiducial value, which means that fitting the linear theory model to data will result in an underestimation of $f\sigma_8$. In the bottom panels, the same quantities are shown but now for the model including non-linear corrections and also the effect of $\Psi^{\delta v}$. Since the template gives a more accurate description of the measured correlation function, there is no offset between the fiducial template (black solid line) and the $1\sigma$ range of the parameter fit (the grey solid region; $0.423<f\sigma_8<0.463$),
and both agree with the measured values to high precision. When fitting the linear model, we note that the $+1\sigma$ limit of the grey shaded region is slightly lower than the mean of the measurement (cf. blue points).

\begin{figure}\label{figure:SN_Quijote}
\centering 
\includegraphics[width=0.47\textwidth]{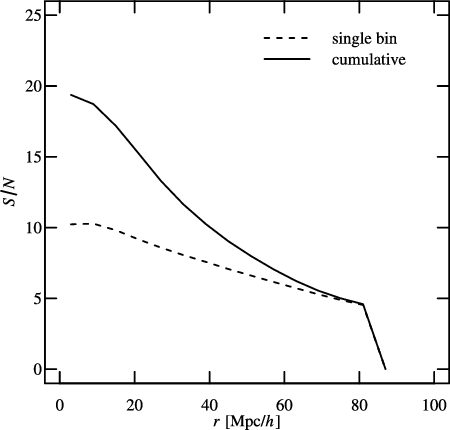}
\caption{The signal-to-noise ratio of the combination of $\psi_1(r)$ and $\psi_2(r)$. The lines indicate combined $S/N$ of $\psi_1(r_i)/\Delta \psi_1(r_i)$ and $\psi_2(r_i)/\Delta \psi_2(r_i)$. The solid line is the cumulative $S/N$ considering all bins above a given $r$, and the dashed line is the $S/N$ per bin.}
\end{figure}

Figure \ref{figure:SN_Quijote} presents the $S/N$ ratio of the measurements as a function of separation. The $S/N$ is defined as the ratio between the mean of the measurement and its standard deviation. We use both $\psi_1(r)$ and $\psi_2(r)$ in our study, and these quantities have a non-zero correlation. We account for the correlation by adopting the following $S/N$ measure;

\begin{equation} S/N= \sqrt{(\psi_1(r),\psi_2(r))^{\rm T}[{\rm Cov}(\psi_1(r),\psi_2(r))](\psi_1(r),\psi_2(r))} ,
\end{equation}

\noindent which is an explicit function of separation $r$. This $S/N$ corresponds to the $\chi^2$ value assuming the null model. We also calculate the cumulative $S/N$ similarly, using $\psi_1$ and $\psi_2$ in all bins in the range $[r,90] \, h^{-1} \, {\rm Mpc}$. At $\sim 30 \, h^{-1} \, {\rm Mpc}$, $S/N\sim10$ for a single bin and it is comparable with the cumulative $S/N$ at $50 \, h^{-1} \, {\rm Mpc}$ (that is, the sum of bins in the range $[50,90] \, h^{-1} \, {\rm Mpc}$). Thus, the constraining power predominantly arises from the smallest separation bin, as expected.

The attainable precision $\Delta (f\sigma_8)/f\sigma_8$ is $4.5$ per cent
($r_{\rm min} =  33\, h^{-1} \, {\rm Mpc}$) and $7.3$ per cent
($r_{\rm min} =  57\, h^{-1} \, {\rm Mpc}$).
From Figure \ref{figure:theory} (right panel), the effect of both higher order perturbative terms and density-velocity correlations are $5$--$10$ per cent, resulting in a $10$--$20$ per cent difference in the theoretical template. This in turn is translated to a $5$--$10$ per cent shift in the best-fitting $f\sigma_8$, which is similar to the statistical uncertainty of the DESI-sized PV survey. While linear theory works sufficiently for an SDSS-sized survey, the importance of an accurate theoretical template will be important in the upcoming generation. 

\begin{figure}
\centering
\includegraphics[width=0.47\textwidth]{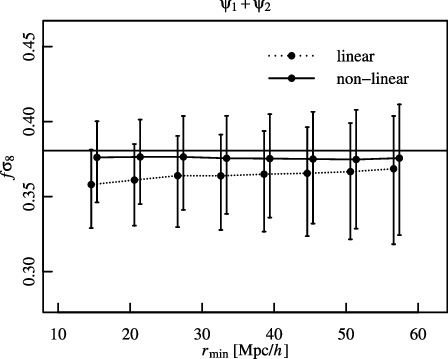}
\caption{\label{figure:fsig8_as_rmin_HR4} Similar to Figure \ref{figure:fsig8_as_rmin_Quijote}, but for the HR4 mock galaxy samples.}
\end{figure}

\begin{figure}
\centering 
\includegraphics[width=0.47\textwidth]{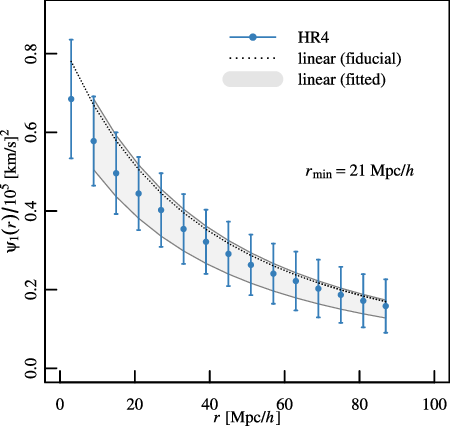}
\includegraphics[width=0.47\textwidth]{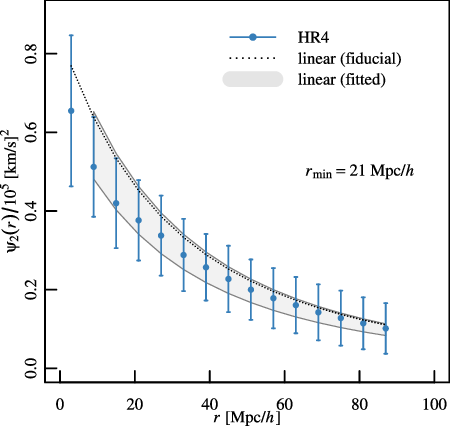}\\
\includegraphics[width=0.47\textwidth]{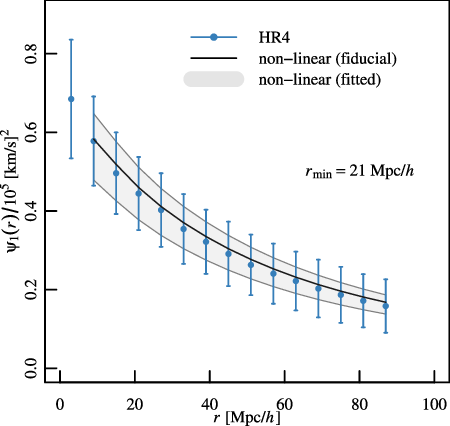}
\includegraphics[width=0.47\textwidth]{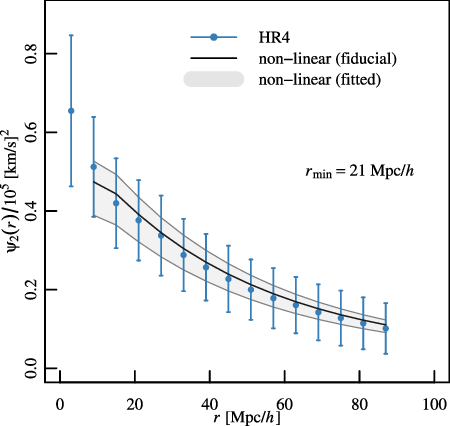}
\caption{\label{figure:bestfit_HR4}Similar to figure \ref{figure:fsig8_as_rmin_Quijote}, but for the HR4 mock galaxy samples.}
\end{figure}

Figures \ref{figure:fsig8_as_rmin_HR4} and \ref{figure:bestfit_HR4} show the same quantities as in Figures \ref{figure:fsig8_as_rmin_Quijote} and \ref{figure:bestfit_Quijote}, but using the HR4 galaxy mock catalogs. The fitted value at $r_{\rm min}=21 \, h^{-1} \, {\rm Mpc}$ is $f\sigma_{8}=0.361^{+0.024}_{-0.030}$ (linear)
and $f\sigma_{8} = 0.376^{+0.025}_{-0.031}$ (PT).
We also performed the fitting without applying the velocity measurement error to the mock galaxies, which allows us to disentangle the different sources of uncertainty in the measurement -- PV and cosmic variance. Assuming that these two contributions can be added in quadrature, and that the fit without any velocity error is purely cosmic variance limited, we obtain the following;
$f\sigma_{8} = 0.361\left(^{+0.020}_{-0.027}\right)_{{\rm PV}}\left(^{+0.013}_{-0.015}\right)_{{\rm cosmic}}$ (linear) 
and 
$f\sigma_{8} = 0.376\left(^{+0.021}_{-0.027}\right)_{{\rm PV}}\left(^{+0.014}_{-0.016}\right)_{{\rm cosmic}}$
(PT).
\citet{Dupuy:2019} separately provided the error bars $f\sigma_8=0.43(\pm0.03)_{\rm PV}(\pm0.11)_{\rm cosmic}$. Our PV error is slightly smaller than their result, which is likely attributed to the difference in the sample size.
For the volume of current available data $\sim (150 \, h^{-1} \, {\rm Mpc})^3$, the cosmic variance dominates the velocity uncertainty, while for a DESI-sized survey, the effect of cosmic variance is smaller than the PV error.
Considering that it would be difficult to obtain PV with a precision significantly better than the current level, the constraining power on $f\sigma_8$ will be limited by the PV error ultimately, while the PV surveys to date are limited by cosmic variance. 
The constraint from linear theory $f\sigma_8=0.361$ is $5.2$ per cent lower than the fiducial value. The systematic offset due to using the linear theoretical template at $r=21\, h^{-1} \, {\rm Mpc}$ is $9.1$ per cent ($\psi_1$) and $13.4$ per cent ($\psi_2$), whose square root is close to the corresponding discrepancy in $f\sigma_{8}$. The precision $\Delta f\sigma_8/(f\sigma_8)$ is $7.5$ per cent and this value is slightly better than the forecast of \citet{Saulder:2023}. This is likely due to the lower limit of the scale ($20 \, h^{-1} \, {\rm Mpc}$ compared to $k\sim 0.2 \, h \, {\rm Mpc}^{-1}$). We conclude that ignoring the non-linear effect causes a $\sim 1\sigma$ level systematic for a DESI-sized PV survey.

\section{Discussion}\label{sec:discussion}

In the near future, a new generation of galaxy catalogs \citep{Aghamousa:2016,Amendola:2018,Tully:2023} will allow us to map the velocity field of the Universe with unprecedented precision, and for the first time over cosmological scales. Having accurate theoretical templates to extract information from these data sets is imperative.

In this work we have calculated the two-point correlation function for the peculiar velocity in real space, using initially linear theory and then tree order perturbation theory up to $\delta_{\rm L}^{4}$. In doing so, we found that the non-linear perturbation terms contribute beyond the linear theory prediction by $5$--$10$ per cent at $r=10$--$20 \, h^{-1} \, {\rm Mpc}$. We additionally accounted for the often neglected fact that the observables are generated as density-weighted quantities. We showed that the density-velocity term contributes a similar $5$--$10$ per cent to the correlation function. As both effects are negative, the application of linear theory to scales at $r_{\rm min} < 20  \, h^{-1} \, {\rm Mpc}$ causes an underestimation of $f\sigma_8$ by $5$--$10$ per cent. 

To confirm our results, we measured the two point statistics from dark matter particle and mock galaxy, $z=0$ snapshot boxes from the Quijote and Horizon Run 4 simulation projects. After generating $N_{r} \sim 240$, $V=(500 \, h^{-1} \, {\rm Mpc})^{3}$ and $N_{r} = 216$, $V=(525 \, h^{-1} \, {\rm Mpc})^{3}$ sub-boxes from the Quijote and HR4 data respectively, we measured $\psi_{1}$, $\psi_{2}$ and fitted two different theoretical templates to the mean of the realisations -- first assuming linear theory and second using perturbation theory to order $\delta_{\rm L}^{4}$ and also including the density-velocity cross power. By fitting the linear theory expectation value to the measurements, we found a systematic bias in the reconstructed value of $f\sigma_{8}$ which was largely eliminated when we modelled the correlation function more accurately beyond linear theory. The bias is sensitive to the choice of scales used in the analysis, and we found that linear theory is approximately valid on scales $r_{\rm min} \sim 50 \, h^{-1} \, {\rm Mpc}$. 
This work highlights the importance of the theoretical template, in particular non-linear effects, when attempting to extract cosmological parameters from the two-point velocity statistics down to smaller scales. 
Using mock galaxy catalogs, \citet{Turner:2021} studied the velocity correlation function. The fitting using only $\psi_1$ and $\psi_2$ resulted in slightly lower $f\sigma_8$ value than the simulation input.
This could be partly attributed to the effects of non-linearity and $\Psi^{\delta v}$. In comparison, \citet{Adams:2017} used a different approach, gridding the catalog and defining the density and velocity of each grid-point. In this case, the quantities are volume-weighted, and the cross correlation $\langle \delta v \rangle$ does not enter into the statistic, at a cost that the scales smaller than the grid resolution cannot be probed.

By performing the fitting to the HR4 mock data with and without PV error, we separately derived the error bars arising from the cosmic variance and PV errors, $f\sigma_{8} = 0.376\left(^{+0.021}_{-0.027}\right)_{{\rm PV}}\left(^{+0.014}_{-0.016}\right)_{{\rm cosmic}}$.
Comparing to the values from the current size of the survey, the cosmic variance is much smaller, responding to the increase in the survey volume, and the PV error shows a moderate improvement.
Although the increase in galaxy number density partially reduces the PV error, the increase in the survey distance increases the PV error. We thus expect that the constraining power on $f\sigma_8$ will be limited by the PV error in the future.

Our analysis of mock galaxy catalogs largely mimicked the dark matter case, which is expected as the velocity correlation function is practically insensitive to the bias of the tracer particles. Some bias dependence is present, by virtue of the density-velocity cross power $\Psi^{\delta v}$ term. However, we find that the same non-linear template can be used for mock galaxies and dark matter particles. Adding information from the density-density power spectrum would allow for simultaneous constraints on $b\sigma_{8}$ and $f\sigma_{8}$.

While the present study is limited to real space, actual observations are typically made in redshift space.
In redshift space, the velocity correlation function is not only a function of the separation, but also the angle between line of sight and the separation vector (see Appendix. D of \citet{Dam:2021}). Even when we take an angular average (a monopole), there is a crosstalk from higher-order multipole moments of $P_{\theta\theta}$. Due to such an additional angular dependence, the relation between ($\Psi_{\parallel}$, $\Psi_{\perp}$) and ($\psi_1$, $\psi_2$) may be more complicated than the real-space analysis. 
Besides the complication of angular dependence, the effect of redshift-space distortion on the velocity divergence power spectra itself is an outstanding problem. The additional non-linear correction to $P_{\theta\theta}$ caused by redshift-space distortion is comparable to real-space non-linearities (see Figure 2 of \citet{Dam:2021}). These effects will systematically bias $f\sigma_{8}$ extracted from the data to even lower values.
The finger-of-god effect; random motion within collapsed galaxy groups, is also a serious issue. 
\citet{Dam:2021} found that at $k\sim 0.2 \, h \, {\rm Mpc}^{-1}$, the redshift-space distorted velocity power spectrum is $\sim 30$ per cent lower than the real space counterpart. They attribute approximately half of this reduction to the redshift-space perturbation theory contributions and the rest to this FoG. This behavior can be also found in \citet{Dupuy:2019} and \citet{Appleby:2023}. Although the galaxy groups are of a few $h^{-1} \, {\rm Mpc}$, the effect of FoG extends to as large as $r \sim 60 \, h^{-1} \, {\rm Mpc}$, requiring the modeling of the FoG effect.
Repeating our analysis in redshift space remains an interesting direction of future study.

\section*{Acknowledgements}
MT and SA are supported by an appointment to the JRG Program at the APCTP through the Science and Technology Promotion Fund and Lottery Fund of the Korean Government, and were also supported by the Korean Local Governments in Gyeongsangbuk-do Province and Pohang City. 
This work was supported in part by the National Research Foundation of Korea (NRF) grant funded by the Korea government (MSIT) (2022R1F1A1064313).
SEH was partly supported by the projects \begin{CJK*}{UTF8}{mj}우주거대구조를 이용한 암흑우주 연구\end{CJK*} (``Understanding Dark Universe Using Large Scale Structure of the Universe,'' No. 1711195982) and \begin{CJK*}{UTF8}{mj}차세대의 우주론적 유체역학 수치실험 DARWIN 수행을 통한 왜소은하의 형성기원 규명\end{CJK*} (``DARWIN: DAzzling Realization of dWarf galaxies In the Next generation of cosmological hydrodynamic simulations,'' No. 1711193044), funded by the Ministry of Science. JK was supported by a KIAS Individual Grant (KG039603) via the Center for Advanced Computation at Korea Institute for Advanced Study. We thank the Korea Institute for Advanced Study for providing computing resources (KIAS Center for Advanced Computation Linux Cluster System). A large data transfer was supported by KREONET, which is managed and operated by KISTI.

\section*{Data Availability}
The data underlying this article will be shared on reasonable request to the corresponding author.
 



\bibliographystyle{mnras}
\bibliography{./main_rev1} 



\clearpage
\appendix
\section{Ensemble Averages of $\psi_1$ and $\psi_2$ as Fractions}\label{appendix:supp}
\begin{figure}
\centering 
\includegraphics[width=0.47\textwidth]{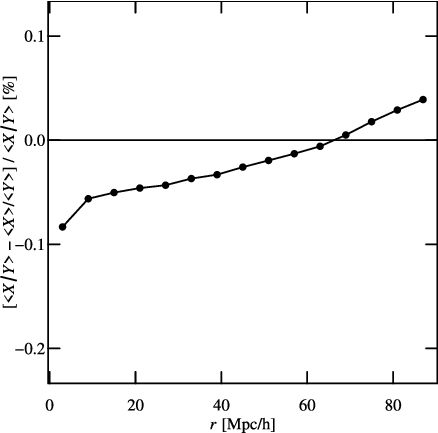}
\includegraphics[width=0.47\textwidth]{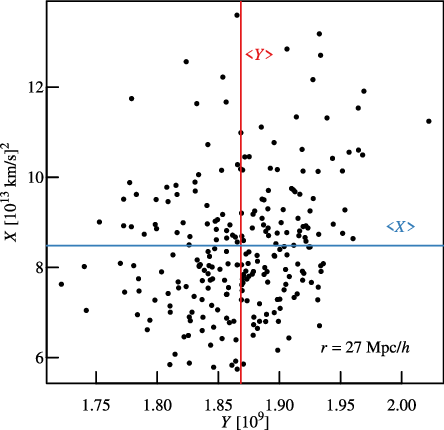}
\caption{\label{fig:XY} ({\it Left Panel}) The percentage fractional difference between $\langle X/Y \rangle$ and $\langle X \rangle / \langle Y \rangle$, obtained from $240$ measurements with the Quijote data. See the text for the definitions of $X$ and $Y$.
({\it Right Panel}) The scatter plot of $X$ and $Y$ of the $240$ measurements. The averages are indicated as solid red/blue lines.}
\end{figure}

In equation (\ref{eq:p1ea}) and (\ref{eq:p2ea}), we applied the ensemble average separately in the numerator and denominator. In general, this is not exact because $\langle X/Y \rangle \neq \langle X \rangle / \langle Y \rangle$. However, we verified that, $\langle X/Y \rangle$ is very close to $\langle X \rangle / \langle Y \rangle$ using the $240$ Quijote simulation samples described in the main text. Here, $X=\sum_{\rm D} w_1 w_2 u_1 u_2 \cos{\theta_{12}}$ and $Y=\sum_{\rm D} w_1 w_2 \cos^2{\theta_{12}}$.  Figure \ref{fig:XY} (left panel) shows the fractional difference between $\langle X/Y \rangle$ and $\langle X \rangle / \langle Y \rangle$ for different $r$ bins. The difference is sub-percent, supporting the operation done in the main text. The right panel shows the scatter plot of $X$ and $Y$ at a fixed scale $r=27 \ {\rm Mpc}/h$. There is a positive correlation between $X$ and $Y$. The correlation coefficient is $0.25$. The averages $\langle X \rangle$ and $\langle Y \rangle$ are indicated as blue and red lines, and the intersection indicates $\langle X \rangle / \langle Y \rangle$. 
There are four regions separated by the lines. In the top-right and bottom-left regions, both $X$ and $Y$ increase (or decrease), and $X/Y$ will be similar to $\langle X \rangle / \langle Y \rangle$.
In the bottom-right region, $X$ is larger than $\langle X \rangle$ and $Y$ is smaller than $\langle Y \rangle$, so $X/Y$ tends to be large. In the top-left region, the opposite occurs and $X/Y$ will be small.
Combining all regions, the mean $\langle X/Y \rangle$ became similar to $\langle X \rangle / \langle Y \rangle$ because of the cancellation.  Even though there is a moderate correlation between $X$ and $Y$, the correlation will not cause the deviation of $\langle X/Y \rangle$ from $\langle X \rangle / \langle Y \rangle$ when the number of points in the top-left and bottom-right subregions are similar.

\section{Derivations of the correlation functions}\label{appendix:derivation}
The velocity-velocity correlation function is calculated as follows \citep{Dodelson:2003}:
\begin{eqnarray}
   \langle v_i(\vecr_1) v_j(\vecr_2) \rangle
&=& \int \frac{{\rm d}^3k}{(2\pi)^3}e^{-\rmi\veck \boldsymbol{\cdot} \vecr_1}  \int \frac{{\rm d}^3k'}{(2\pi)^3}e^{-\rmi\veck' \boldsymbol{\cdot} \vecr_2} \langle  \tilde{v}_i(\veck) \tilde{v}_j(\veck') \rangle \\
&=& -(aHf)^2 \int \frac{{\rm d}^3k}{(2\pi)^3} \int \frac{{\rm d}^3k'}{(2\pi)^3} \frac{k_i k'_j}{k^2 k'^2} \langle \tilde{\theta}(\veck) \tilde{\theta}(\veck') \rangle  e^{-\rmi(\veck \boldsymbol{\cdot} \vecr_1 + \veck' \boldsymbol{\cdot} \vecr_2)} \\
&=& (aHf)^2 \int \frac{{\rm d}^3k}{(2\pi)^3} \frac{k_ik_j}{k^4}P_{\theta\theta}(k)e^{-\rmi\veck \boldsymbol{\cdot}(\vecr_1 - \vecr_2)}\\
&=& (aHf)^2\int \frac{{\rm d}k}{(2\pi)^3 k^2}P_{\theta\theta}(k)\int d\Omega_k k_i k_j e^{-\rmi\veck \boldsymbol{\cdot}\vecr} \\
&=& -(aHf)^2 \int \frac{{\rm d}k}{(2\pi)^3k^2}P_{\theta\theta}(k) \frac{{\rm d}^2}{{\rm d}r_i {\rm d}r_j} \int d\Omega_k e^{-\rmi\veck\boldsymbol{\cdot}\vecr} \\
&=& -(aHf)^2  \int \frac{{\rm d}k}{2\pi^2k^2}P_{\theta\theta}(k) \frac{{\rm d}^2}{{\rm d}r_i {\rm d}r_j} j_0(kr)\\
&=& -(aHf)^2  \int \frac{{\rm d}k}{2\pi^2}P_{\theta\theta}(k) \left[(\delta^{\rm D}_{ij}-\hat{r}_i\hat{r}_j)\frac{j_0'(kr)}{kr} + \hat{r}_i\hat{r}_j j_0''(kr)\right]\\
&=& \hat{r}_i\hat{r}_j \Psi_{\parallel}^{vv} + (\delta^{\rm D}_{ij}-\hat{r}_i\hat{r}_j)\Psi_{\perp}^{vv}.
\end{eqnarray}
In the second line, equation (\ref{eq:theta}) is used. In the third line, $\langle \tilde{\theta}(\veck) \tilde{\theta}(\veck') \rangle = (2\pi)^3\delta^{\rm D}(\veck+\veck')P_{\theta\theta}(k)$. In the fourth line, $\int d\Omega_k$ is the angular integration and in the fifth line, the identity $\int d\Omega_k e^{-\rmi\veck \boldsymbol{\cdot} \vecr} = j_0(kr)$ is used. A slightly different mathematical procedure can be also found in \citet{Blake:2024}.
The density-velocity correlation function is derived as follows:
\begin{eqnarray}
\langle \delta^{\rm g}(\vecr_1) v_i(\vecr_2) \rangle
&=& \int \frac{{\rm d}^3k}{(2\pi)^3}e^{-\rmi\veck\boldsymbol{\cdot}\vecr_1} \int \frac{{\rm d}^3k'}{(2\pi)^3}e^{-\rmi\veck'\boldsymbol{\cdot}\vecr_2} \langle \tilde{\delta}^{\rm g}(\veck) \tilde{v}_i(\veck')\rangle\\
&=& i(aHf) \int \frac{{\rm d}^3k}{(2\pi)^3} \int \frac{{\rm d}^3k'}{(2\pi)^3} \frac{k'_i}{k'^2} \langle \tilde{\delta}^{\rm g}(\veck) \tilde{\theta}_i(\veck')\rangle e^{-\rmi(\veck \boldsymbol{\cdot} \vecr_1 + \veck' \boldsymbol{\cdot} \vecr_2)}\\
&=& -i(aHfb) \int \frac{{\rm d}^3k}{(2\pi)^3}\frac{k_i}{k^2}P_{\delta\theta}(k)e^{-\rmi\veck\boldsymbol{\cdot}(\vecr_1-\vecr_2)}\\
&=& (aHfb) \int \frac{{\rm d}k}{(2\pi)^3}P_{\delta\theta}(k) \frac{{\rm d}}{{\rm d}r_i} \int d\Omega_k e^{-\rmi\veck\boldsymbol{\cdot}\vecr}\\
&=& (aHfb) \int \frac{{\rm d}k}{2\pi^2}P_{\delta\theta}(k) \frac{{\rm d}}{{\rm d}r_i}j_0(kr)\\
&=& b\hat{r}_i \Psi^{\delta v},
\end{eqnarray}

In the linear regime, $P_{\theta\theta}(k)=P_{\delta\theta}(k)=P_{\rm L}(k)$, leading to equations (\ref{eq:Psi_para}) and (\ref{eq:Psi_perp}). 


\bsp	
\label{lastpage}
\end{document}